\documentclass{ws-ijmpd}
\usepackage[super,compress]{cite}
\usepackage{graphicx}
\usepackage{epstopdf}
\usepackage{amsfonts}
\usepackage{amssymb}
\usepackage{epsfig}
\usepackage{hyperref}

\usepackage{dcolumn}
\usepackage{bm}
\usepackage{amsmath}
\usepackage{epstopdf}
\usepackage{amsfonts}
\usepackage{amssymb}
\usepackage{epsfig}
\usepackage{tabularx}
\usepackage{color}
\usepackage{hyperref}

\begin{document}


%
\catchline{}{}{}{}{}
%

\title{Constraints on light dark matter particles using white dwarf stars}

\author{Grigoris Panotopoulos}

\address{Centro de Astrof{\'i}sica e Gravita{\c c}{\~a}o-CENTRA, Departamento de F\'isica, 
\\
Instituto Superior T\'ecnico-IST,
Universidade de Lisboa-UL,
\\
Av. Rovisco Pais 1, 1049-001 Lisboa, Portugal
\\
\href{mailto:grigorios.panotopoulos@tecnico.ulisboa.pt}{\nolinkurl{grigorios.panotopoulos@tecnico.ulisboa.pt}} 
}

\author{Il{\'i}dio Lopes}

\address{Centro de Astrof{\'i}sica e Gravita{\c c}{\~a}o-CENTRA, Departamento de F\'isica, 
\\
Instituto Superior T\'ecnico-IST,
Universidade de Lisboa-UL,
\\
Av. Rovisco Pais 1, 1049-001 Lisboa, Portugal
\\
\href{mailto:ilidio.lopes@tecnico.ulisboa.pt}{\nolinkurl{ilidio.lopes@tecnico.ulisboa.pt}} 
}

\maketitle

\begin{history}
\received{Day Month Year}
\revised{Day Month Year}
\end{history}

\begin{abstract}
We report constraints on the nucleon-dark matter particle cross section using the internal luminosity of observed white dwarf stars in the globular cluster Messier 4. Our results cover the parameter space corresponding to relatively light dark matter particles, in the mass range $0.1~GeV-5~GeV$, which is known to be very difficult to be probed by direct dark matter searches. The additional luminosity coming from self-annihilations of dark matter particles captured inside the stars must not exceed the observed luminosity. Imposing that condition, we obtain for the spin independent cross section of light dark matter particles on baryons $\sigma_{N\chi}$ the upper bound: $\sigma_{N\chi} < 4 \times 10^{-41}{\rm cm^2}$.
\end{abstract}


\ccode{}

\section{Introduction}

Many well-established results coming from high-precision  observations in Astrophysics and Cosmology support our belief that the Universe is forming and expanding at an accelerated rate for which the gravitational pull is dominated by dark matter\cite{zwicky,rubin} and the anti-gravity by dark energy\cite{turner,copeland}. The success of modern observational cosmology in explaining the universe has lead to the establishment of the concordance cosmological model, which is based on cold dark matter and a cosmological constant ($\Lambda CDM$). This is presently  the most economical cosmological model that successfully describes the structure formation of the Universe on large  scales. Nevertheless such a cosmological model  says nothing about the nature of the fundamental particles dark matter is made of. Therefore, the determination of the properties of these unknown types of elementary particles, and the role that such particles play in the formation of the Universe is one of the most important problems that  need to be addressed by Particle Physics and Cosmology. 

A very popular dark matter type are the Weakly Interacting Dark matter (WIMP) particles, which are thermal relics from the Big-Bang. However, other good candidates exist as well, for a good list see e.g. \cite{taoso}. To shine some light into the nature of dark matter, several experiments have been designed: for reviews and dark matter searches see e.g. \cite{DM1,DM2,DM3,DM4,DM5,Zurek:2013wia}. In direct dark matter searches an effort is made to observe the nucleus recoil after a dark matter particle scatters off the material of the detector. These direct detection experiments have put limits on the nucleon-dark matter candidate cross section for a given mass of the dark matter particle \cite{dama,cogent,detection1,detection1b,detection2,cresst,edelweiss,zeplin}. Furthermore, during the last 15 years or so, the use of observational data from astrophysical objects, such as the Sun \cite{ilidio1,ilidio2,ilidio3,Vincent:2015gqa,Garani:2017jcj}, solar-like stars \cite{ilidio4,ilidio5,ilidio6} (see also \cite{ilidio0}), white dwarf (WD) stars and neutron stars \cite{kouvaris0,kouvaris1,kouvaris2,GPIL1,GPIL2}, have
been employed to offer us complementary bounds on the WIMP-nucleon cross section.

If dark matter particles accumulate inside an astrophysical object in considerable amounts, it will significantly modify the properties of the star, leading to contradictions with current astronomical data. In this article we propose to use the luminosity of observed white dwarf (WD) stars in the Messier 4 (M4) globular cluster (GC) (also designated as NGC 6121) \cite{M4}, found in the constellation of Scorpion at $\sim 7200$ light years away, to constrain the parameter space on the nucleon-dark matter particle cross section versus the mass of the dark matter particle. For related works where neutron stars instead are used to constrain DM properties see e.g. \cite{NS1,NS2,NS3,NS4,NS5,NS6,NS7,NS8} and references therein.

Our work is organized as follows: after this introduction, we present the theoretical framework in section two, and we constrain the dark matter parameter space in the third section. Finally we summarize and conclude our work in the fourth section. We work in units in which the speed of light in vacuum $c$, the Boltzmann constant $k_B$ and the reduced Planck mass $\hbar$ are set to unity, $c=k_B=\hbar=1$. In these units all dimensionless quantities are measured in GeV, and we make use of the conversion rules
$1 \rm m = 5.068 \times 10^{15} GeV^{-1}$, $1 \rm kg = 5.610 \times 10^{26} GeV$ and $1 \rm K = 8.617 \times 10^{-14} GeV$ \cite{guth}.

\section{Summary of basic properties of white dwarf stars}

WD stars were discovered more than 100 years ago, when in 1914 H.~Russell noticed that the object 40 Eridani B was located well below the main sequence on the Hertzsprung-Russell diagram. WDs are old compact objects that mark the final evolutionary stage of the vast majority of the stars \cite{ref1,ref2}. Indeed, more than $95 \%$, and even perhaps up to $98 \%$ of all stars, will die as WD stars \cite{ref3}. About $80 \%$ of WD possess a hydrogen atmosphere (DA type), while the remaining $20 \%$ show a helium atmosphere (DB type) \cite{ref4}. The low-mass white dwarf stars are expected to harbour He cores, while the average mass white dwarf star most likely contain Carbon/Oxygen cores \cite{ref1}. As thermonuclear reactions no longer take place, WDs are cooling down emitting their stored thermal energy. 

Here we shall consider a set of observed white dwarf stars in the M4 GC \cite{set,base2} with masses in the range $\rm (0.3-1.4)~M_{\odot}$, with $M_{\odot} = 2 \times 10^{30}~kg$ being the solar mass, age $\rm t_* \sim 10^{10}~Gyr$, and a core temperature $\rm T_c = 10^6~K$. The mass-to-radius profile may be computed solving the Lane-Emden equation, see the discussion below, and therefore for a given mass the radius of the object can be computed.

Since matter inside WDs can be described by a non-relativistic polytropic star with an equation-of-state (EoS) of the form 
\begin{eqnarray}
p & = & K \rho^\gamma, 
\label{eq:poly}
\end{eqnarray}
where $p$ and $\rho$ are the pressure and energy density of the stellar  matter, and $ \gamma = 1 + 1/n $ where  $n$ is the polytropic index.
We determine the internal structure of a WD by solving the Lane-Emden equation  for the function $\theta(\xi)$, see e.g. \cite{textbook}, for the initial conditions $ \theta(0)  =  1 $ and $ \theta'(0)  =  0 $, where the primes denote differentiation with respect to $\xi$. 
The radius $R$ of the star is determined by the condition $\theta(\xi_1)=0$. The new variables $(\xi,\theta)$ are related to the original ones $(r,\rho)$ via
$\xi = r/a$ and $\theta^n=\rho/\rho_c$, with $\rho_c$ being the central energy density, while the constant $a$ is defined to be such that
$a^2 ={(n+1) p_c}/{(4 \pi G_N \rho_c^2)} $ where 
$G_N$ is the Newton's constant. Once the root $\xi_1$ is known, the radius $R$ and the mass $M$ of the star are computed by
\begin{equation}
R = a \xi_1
\end{equation}
and
\begin{equation}
M = 4 \pi a^3 \rho_c \int_0^{\xi_1} dz z^2 [\theta(z)]^n.
\end{equation}
By combining the two equations and eliminating the central energy density we obtain the mass-to-radius profile of the form
\begin{equation}
M(R) \sim R^{(n-3)/(n-1)}.
\end{equation}
It follows that the structure of White Dwarfs are described by this well-known Chandrasekhar model \cite{chandra}, for which
the stellar matter with atomic number $A$ and charge $Z$
inside the WD is considered to be  completely ionised. In this stellar model the gravitational force is balanced by the pressure of an ideal degenerate electron gas. Accordingly, in the non-relativistic limit the WD equation-of-state has the following simple form 
\cite{textbook,model1,model2}:
\begin{equation}
p = 10^{13} \: \left( \frac{\rho}{\mu_e} \right)^{5/3}\;\; {\rm (cgs)},
\end{equation}
where  $\mu_e=A/Z$ is the molecular weight per electron,
not to be confused with the chemical potential. For our study, it is 
sufficient to consider that $\mu_e$ has a value equal to 2,
as  in most realistic WD compositions \cite{model2}. By comparing the previous equation with equation (\ref{eq:poly}), we found that the polytropic index $n=3/2$. Although a more sophisticated treatment  is possible (see e.g. \cite{saltas}), for our purposes here it is sufficient to consider the Chandrasekhar WD model.


\begin{figure}[ht!]
\centering
\includegraphics[width=\linewidth]{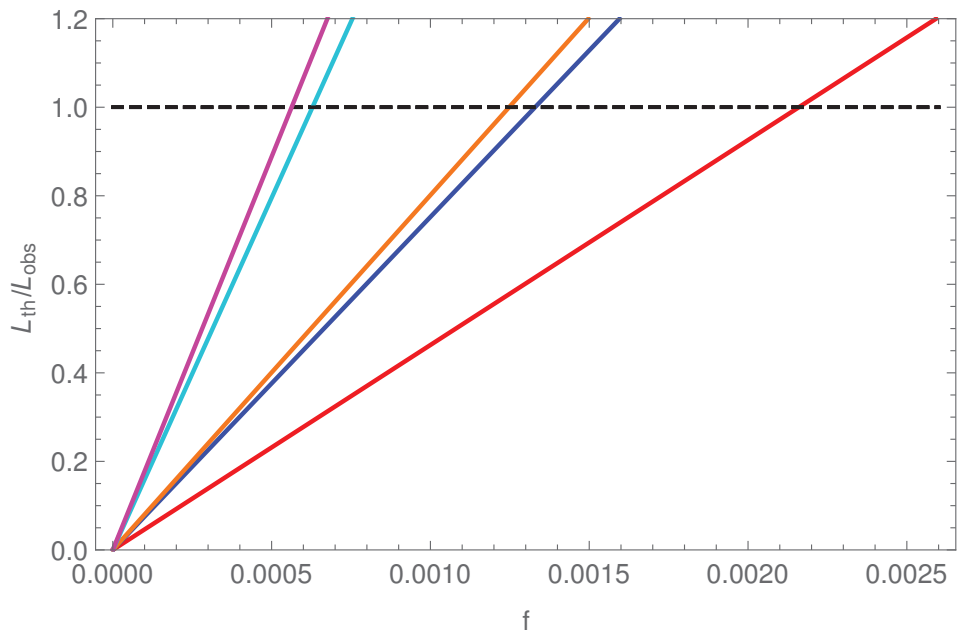}
\caption{
Dimensionless luminosity versus $f$ for the 5 stars shown in Table~\ref{table:stars}.  
The different colour lines ($1\cdots 5$, from the  left-side  to the right-side) correspond  to the total luminosity emitted by each of the WD shown in Table~\ref{table:stars}.  The horizontal dashed line defines the normalized observed luminosity of each WD in the same table.
}
\label{fig:luminosity1} 
\end{figure}

\begin{figure}[ht!]
\centering
\includegraphics[width=\linewidth]{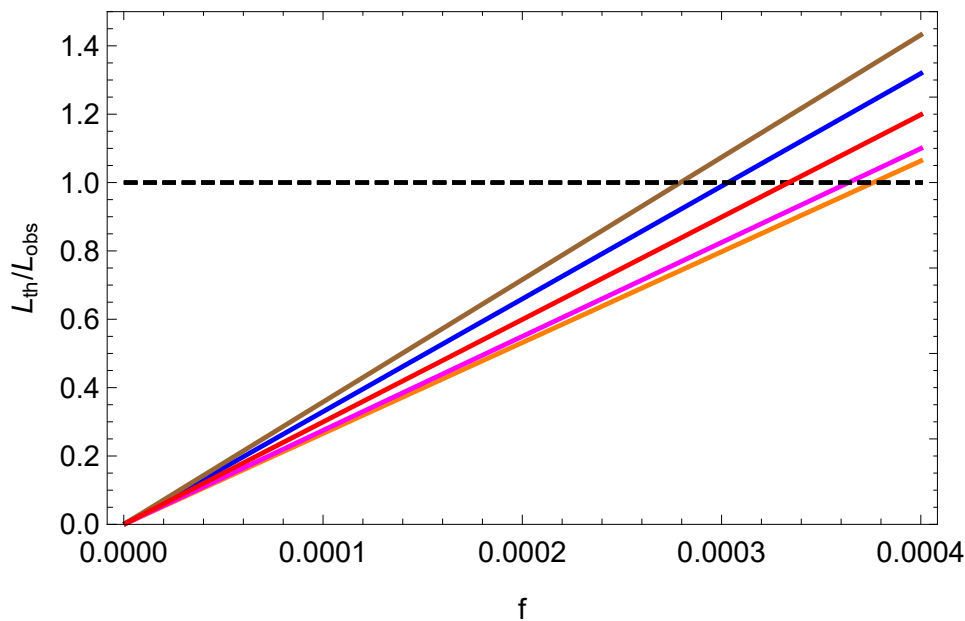}
\caption{
Same as in Figure~1, but for the second group of 5 WD stars shown in Table~\ref{table:starsII}
}
\label{fig:luminosity2} 
\end{figure}


\section{Dark matter accretion inside WDs}

We assume a Gaussian profile for the dark matter number density \cite{dearborn1,dearborn2,kaplan}
\begin{equation}
n(r) = N \pi^{-3/2} r_{\chi}^{-3} e^{-r^2/r_{\chi}^2},
\end{equation}
where the dark matter particles are thermalized inside a sphere of radius \cite{kouvaris1,base2}
$r_{\chi}$ given by
\begin{equation}
r_{\chi} = \sqrt{\frac{9 T_c}{8 \pi G_N m_\chi \rho_c}},
\end{equation}
where $\rho_c$ and $T_c$ are the central density and temperature in the core of the star,
while the number of dark matter particles $N(t)$ accumulated inside the star as a function of time $t$ is computed by the following rate equation \cite{gould1,gould2,kouvaris3,Griest:1986yu}:
\begin{equation}
\frac{dN}{dt} = C_c - C_a N^2 -C_e N,
\label{eq:Nt}
\end{equation}
where in principle there are three competing mechanisms, namely the accretion rate $C_c$ (first term), the self-annihilation contribution $C_a$ (second term) and the evaporation contribution $C_e$ (last term). The latter may or may be not as important as the other two, and in many studies it is neglected. For a given type of object (solar-like stars, WDs, neutron stars etc), it is efficient only for light DM particles \cite{ilidio7} with a mass lower than a critical mass depending on the density of the star. For a solar-like star it is of the order of the GeV \cite{gould2}, but for neutron stars it is of the order of keV \footnote{C.~Kouvaris, private communication}. For WDs we expect something in between, and therefore for the DM mass range considered in the present work, we can safely neglect the evaporation term.

The self-annihilation and the capture rates, $C_a$ and $C_c$, respectively, are given by the expressions
\begin{eqnarray}
C_a & = & \frac{\langle  \sigma v \rangle_\chi }{V_\chi} 
\end{eqnarray}
and
\begin{eqnarray}
C_c & = & \frac{\sqrt{24 \pi} G_N \rho_{\chi} M R}{m_\chi  \bar{v}} f \left[1 - \frac{1-e^{-B^2}}{B^2} \right],
\end{eqnarray}
where $V_\chi$ is the thermalization volume that is given by $V_\chi=(4 \pi/3) r_\chi^3$, 
and  $\langle \sigma v \rangle_\chi \simeq 3 \times 10^{-26} cm^3/s$ 
is the DM annihilation cross section that is required to reproduce the observed DM abundance \cite{Planck2015,Planck2018}. Moreover,  the quantity $B$ is computed as $ B^2 = {6 m_\chi v_{esc}^2}/{m_A \bar{v}^2 (m_\chi /{m_A}-1)^2} $ where  $m_A$ is the mass of a nucleus with baryonic number $A=14$ to account for the mixed $C-O$ composition found inside the WD. The escape velocity is given by
$ v_{esc} = \sqrt{{2 G_N M}/{R}} $ and $\bar{v}=20\;{\rm km/s}$. In addition, the fraction of particles that undergo one or more scattering is given by $f \simeq \sigma/\sigma_{cr} = A^2 \sigma_{N\chi}/\sigma_{cr}$ \cite{base2}, with the saturation cross section being $\sigma_{cr} = {\pi m_A R^2}/{M}$,
while the total nucleus-dark matter particle cross section $\sigma=A^2 \sigma_{N\chi}$.
Finally, using the Navarro-Frenk-White halo profile \cite{NFW}
\begin{equation}
\rho_{NFW}(r) = \frac{\rho_s}{(r/r_s) (1+r/r_s)^2}
\end{equation}  
where $r_s=20~kpc$ and $\rho_s=0.26~GeV/cm^3$ \cite{cirelli}, one can compute the local value of DM density at the Solar System, which is located at $\sim 8.5~kpc$ from the galactic center, and it is found to be $\rho_0=0.3~GeV/cm^3$.
We remark in passing that this value of $\rho_0$ is rather conservative, since current observations suggest $\rho_0 \simeq 0.38 GeV/cm^3$, while some others indicate a value two times larger (see \cite{Catena:2009mf,ilidio1,ilidio2} for details). The DM density at GC M4 was estimated to be $\rho_\chi=798~GeV/cm^3$, see \cite{set} for the details.
Therefore $\rho_\chi=2660 \: \rho_0$, a value used also in \cite{base2}. 

With the initial condition $N(0)=0$, the rate equation (\ref{eq:Nt})
can be easily integrated, and thus the number of dark matter particles accumulated inside the star during its lifetime is given by
\begin{equation}
N = \sqrt{\frac{C_c V_\chi}{\langle \sigma v \rangle_\chi}} \: \textrm{tanh} \left( \sqrt{\frac{C_c \langle \sigma v \rangle_\chi}{V_\chi}} t_* \right),
\end{equation}
where $t_*$ is the age of the star.

It is worth mentioning that the exact solution above acquires a simpler form in two limiting cases in which the argument of the function $tanh(x)$ is either very small or very large: (i) for which $x$ is very small ($x \ll 1$), in this case the annihilation cross section can be neglected; (ii) for which $x$ is very large ($x \gg 1$), this is the case that occurs after a sufficiently long time that the dark matter particles reaches the equilibrium in which the two competing mechanisms in the rate equation (\ref{eq:Nt}) cancel one another and the number of DM particles remains the same. In the first case (i) one finds $N \simeq C_c t_*$, which can be obtained from the rate equation neglecting the annihilation term, while in the second case (ii) one finds $N \simeq  \sqrt{{C_c V_\chi}/{\langle \sigma v \rangle_\chi}}$
which can be obtained from the rate equation (\ref{eq:Nt}) setting $dN/dt=0$. It is easy to verify that in our work, given the numerical values at hand corresponds  to the  case (ii), we can use the previous formula for $N$ at equilibrium.

Dark matter in the stellar core provide an additional source of energy \cite{salati,Lopes:2014xaa}. The energy generation rate $\epsilon_\chi$ due to DM pair annihilation is
given by \cite{ilidio2}
\begin{equation}
\epsilon_\chi = m_\chi n(r)^2 \rho(r)^{-1} \langle \sigma v \rangle_\chi
\end{equation}
and therefore the internal luminosity of WD stars due to pair annihilation of DM particles is computed by \cite{ilidio2,salati}
\begin{equation}
\frac{dL_\chi}{dr} 
= 4 \pi r^2 \rho(r) \epsilon_\chi 
= 4 \pi r^2 n(r)^2 m_\chi \langle \sigma v \rangle_\chi
\end{equation}
upon integration from the center $r=0$ to the surface $r=R$ of the star. 

We can imagine simple dark matter models with a handful of free parameters, such as the mass of the dark matter particle $m_\chi $ and some coupling constants $\lambda_i$. For a concrete, simple, non-supersymmetric model of this class see e.g. \cite{singlet1,singlet2,singlet3,singlet4,inert1,inert2,inert3,Hportal}. The requirement that the dark matter abundance agrees with the PLANCK value, $\Omega_{\chi} h^2 \simeq 0.12$ \cite{Planck2015,Planck2018}, reduces the number of free parameters by one. Furthermore, the nucleon-dark matter particle cross section $\sigma_{N\chi}$ may be computed in terms of the parameters of the model at hand, but we can trade one of the coupling constants for $\sigma_{N\chi}$. Therefore, in the rest of the article we shall assume that $(m_\chi,\sigma_{N\chi})$ are the two free parameters to be constrained.

Regarding direct dark matter searches, the current situation (limits and hints) is illustrated in figures 26.1 and 26.2 of the review on Dark Matter by M.~Tanabashi et al. (Particle Data Group) \cite{PDG}. For dark matter particle masses below $10~GeV$ the limit is set by solar neutrinos. Therefore, we consider in this work the case of a relatively light dark matter particle, with a mass of the order of the GeV, since this corner of the parameter space cannot be probed by direct dark matter searches (see also the Fig.~1 of \cite{Hportal}).

In the tables~\ref{table:stars} and~\ref{table:starsII} below we show the mass and the luminosity of ten observed WDs used in the numerical analysis. We have organized them into two groups of five stars each, and the second group is characterized by lower luminosities. For a given WD the radius is computed using the results obtained integrating the Lane-Emden equation. Since $\sigma_{cr}$ is known, the internal luminosity is a function of ${m_\chi,\sigma_{N\chi}}$ (or of ${m_\chi ,f}$) only. Our main numerical results are summarized in Figures~\ref{fig:luminosity1} and\ref{fig:luminosity2} where we show the internal luminosity of WDs due to DM self-annihilations versus the fraction $f$ setting $m_\chi=1~GeV$ for the ten stars shown in the two tables, and the corresponding upper bound on $f$ for each case. We have checked for $m_\chi=0.1~GeV$ or for $m_\chi=5~GeV$ the figures do not change,
which implies that in the mass range considered here, $0.1~GeV \leq m_\chi \leq 5~GeV$, the internal luminosity of WD stars due to DM self-annihilations is not sensitive at all to the mass of the DM particle.

Our numerical results show that the bounds set by the second group are more stringent, as expected since the luminosity is lower there. This upper bound implies an upper bound on the nucleon-dark matter particle cross section $\sigma_{N\chi}$. Considering the stringiest bounds (magenta curve in the first figure, brown curve in the second figure) 
\begin{equation}
f \leq 5 \times 10^{-4}
\end{equation}
and
\begin{equation}
f \leq 2.8 \times 10^{-4}
\end{equation}
we obtain the corresponding upper bound for the nucleon-dark matter particle cross section:
\begin{equation}
\boxed{\sigma_{N\chi} \leq 4 \times 10^{-41}{\rm cm^2}}
\end{equation}
from the first group, and
\begin{equation}
\boxed{\sigma_{N\chi} \leq 2 \times 10^{-41}{\rm cm^2}}
\end{equation}
from the second group. 

Those are the main results of our work obtained for $m_\chi=1~GeV$. Since the internal luminosity of WD stars due to DM self-annihilations is not sensitive to the mass of the DM particle, we obtain precisely the same bounds of the factor $f$ throughout the whole mass range.

A final remark is in order here. In this class of articles it is a common practice to draw two-dimensional plots showing the nucleon-dark matter particle cross section as a function of the mass of the DM particle. Those figures allow a direct comparison with the results obtained in previous analyses. This, however, is not possible in the present work. The reason is that for the mass range considered here, $0.1~GeV \leq m_\chi \leq 5~GeV$, the bounds on $f$ do not depend on $m_\chi$, as already mentioned before, and therefore the corresponding bounds on $\sigma_{N\chi}$ are mass independent.

Finally, we require that the thermalization condition $t_\chi < t_*$ is satisfied, which requires that the thermalization time $t_\chi$ does not exceed 
the age of the star, $t_*=(12.7 \pm 0.7)~Gyr$ \cite{age}, and $t_\chi$ is given by \cite{kouvaris1}
\begin{eqnarray}
\frac{t_\chi }{\rm 4\, yr}= \left( \frac{m_\chi}{\rm TeV} \right)^{3/2} \left( \frac{10^8 \rm g\, cm^{-3}}{\rho_\star} \right) \left( \frac{10^{-43} \rm cm^2}{f \sigma_{cr}} \right)
\left( \frac{10^7 \rm K}{T_c} \right)^{1/2}
\end{eqnarray}
where $\rho_*$ is the typical energy density of ordinary matter, $\rho_\star \sim 10^{9}\rm kg\,m^{-3}$. This condition implies a lower limit of the factor $f$ for a given mass $m_\chi$, and a corresponding lower limit on $\sigma_{N\chi}$. In particular, we find that
for $m=0.1 \; {\rm GeV} $, $1 \; {\rm GeV}$ and $5 \; {\rm GeV}$, 
$f$ is larger than $1.4 \times 10^{-19}$, $4.4 \times 10^{-18}$ 
and $4.9 \times 10^{-17}$, respectively, while for the corresponding nucleon-DM particle cross sections we obtain that $\sigma_{N\chi}$ is larger than $1.0 \times 10^{-56}\;{\rm cm^2}$, $3.2 \times 10^{-55}\;{\rm cm^2}$, and  $3.5 \times 10^{-54}\;{\rm cm^2}$.


\vskip 0.5cm

\begin{table}[ht!]
\tbl{Masses and luminosities of 5 observed WDs (first group) in the globular cluster M4.}
{
\begin{tabular}{@{}ccc@{}} \toprule
$N^{\rm o}$ WD star & $M (M_{\odot})$ & $L (10^{28} {\rm erg\, s^{-1}})$	\\
\\ \colrule
1 & 0.28  & 6.318  \\
2 & 0.673 & 6.979  \\
3 & 1.006 & 8.563  \\
4 & 1.154 & 4.728  \\
5 &  1.378 & 4.766  
\end{tabular}
\label{table:stars}
}
\end{table}

\begin{table}[ht!]
\tbl{Masses and luminosities of 5 observed WDs (second group, lowest luminosities) in the globular cluster M4.}
{
\begin{tabular}{@{}ccc@{}} \toprule
$N^{\rm o}$ WD star & $M (M_{\odot})$ & $L (10^{28} {\rm erg\, s^{-1}})$	\\
\\ \colrule
1 & 1.154 & 2.829  \\
2 & 1.260 & 2.901  \\
3 & 1.327 & 2.503  \\
4 & 1.366 & 2.351  \\
5 & 1.351 & 2.788  
\end{tabular}
\label{table:starsII}
}
\end{table}
  

\section{Conclusions}

In summary, in this work we have obtained new constraints on properties of dark matter using the internal luminosity of white dwarf stars observed in the M4 globular cluster. Our results probe the parameter space of light dark matter particles, in the mass range $0.1~GeV-5~GeV$, which cannot be probed by other direct dark matter searches. The additional luminosity coming from the self-annihilations of dark matter particles captured inside the stars, for a given object, depend on the two free parameters of the models, namely the mass of the dark matter particle $m_\chi $, as well as the nucleon-dark matter particle cross section $\sigma_{N\chi}$.  Imposing the condition that dark matter particles with a mass smaller than $5~GeV$,  must not exceed the observed WD luminosity, we obtained for $\sigma_{N\chi}$ the upper bound $\sigma_{N\chi} < 4 \times 10^{-41} {\rm cm^2}$. For a dark matter particle of $1~GeV$, this value is a 3 order of magnitude smaller than the one imposed by the current detectors (cf. Fig.~26.1  in \cite{PDG}, $\sigma_{N\chi}< 10^{-38} {\rm cm^2}$).


\section*{Acknowlegements}

We would like to thank the anonymous reviewer for useful comments and suggestions.
It is a pleasure to warmly thank J.~Lopes and C.~Kouvaris for enlightening discussions. We also wish to thank M.~Cerme{\~n}o and M.~A.~P{\'e}rez-Garc{\'i}a for correspondence and for sending us their data. The authors thank the Funda\c c\~ao para a Ci\^encia e Tecnologia (FCT), Portugal, for the financial support to the Center for Astrophysics and Gravitation-CENTRA, Instituto Superior T\'ecnico, Universidade de Lisboa, through the Project No.~UIDB/00099/2020.



\begin{thebibliography}{99}
\bibitem{zwicky} F.~Zwicky,
 Helv.\ Phys.\ Acta {\bf 6} (1933) 110
   [Gen.\ Rel.\ Grav.\  {\bf 41} (2009) 207].
   
\bibitem{rubin} V.~C.~Rubin and W.~K.~Ford, Jr.,
  Astrophys.\ J.\  {\bf 159} (1970) 379.
  
\bibitem{turner} W.~L.~Freedman and M.~S.~Turner,
  Rev.\ Mod.\ Phys.\  {\bf 75} (2003) 1433
  [astro-ph/0308418].  
  
\bibitem{copeland} E.~J.~Copeland, M.~Sami and S.~Tsujikawa,
  Int.\ J.\ Mod.\ Phys.\ D {\bf 15} (2006) 1753
  [hep-th/0603057].  
  
\bibitem{taoso} M.~Taoso, G.~Bertone and A.~Masiero,
  JCAP {\bf 0803} (2008) 022
  [arXiv:0711.4996 [astro-ph]].
  
\bibitem{DM1} K.~A.~Olive,
  astro-ph/0301505.
  
\bibitem{DM2} C.~Munoz,
  Int.\ J.\ Mod.\ Phys.\ A {\bf 19} (2004) 3093
  [hep-ph/0309346].
  
\bibitem{DM3} J.~Gascon,
  EPJ Web Conf.  {\bf 95}, 02004 (2015).
  
\bibitem{DM4} J.~M.~Gaskins,
  Contemp. Phys.  {\bf 57}, 496 (2016); 
  [arXiv:1604.00014 [astro-ph.HE]].

\bibitem{DM5} F.~Kahlhoefer,
  Int. J. Mod. Phys. A {\bf 32}, 1730006 (2017); 
  [arXiv:1702.02430 [hep-ph]].  
  
 \bibitem{Zurek:2013wia} K.~M.~Zurek,
  Phys.\ Rept.\  {\bf 537} (2014) 91;
  [arXiv:1308.0338 [hep-ph]].
 
\bibitem{dama} R.~Bernabei {\it et al.},
  Frascati Phys.\ Ser.\  {\bf 58} (2014) 41.
  
\bibitem{cogent} C.~E.~Aalseth {\it et al.} [CoGeNT Collaboration],
  Phys.\ Rev.\ D {\bf 88} (2013) 012002
  [arXiv:1208.5737 [astro-ph.CO]].

\bibitem{detection1} E.~Aprile {\it et al.},
  New Astron.\ Rev.\  {\bf 49} (2005) 289.
  
\bibitem{detection1b} Z.~Ahmed {\it et al.} [CDMS-II Collaboration],
  Science {\bf 327} (2010) 1619
  [arXiv:0912.3592 [astro-ph.CO]].
  
\bibitem{detection2} D.~S.~Akerib {\it et al.} [LUX Collaboration],
Phys.\ Rev.\ Lett.\  {\bf 116} (2016) no.16, 161301
[arXiv:1512.03506 [astro-ph.CO]].

\bibitem{cresst} G.~Angloher {\it et al.} [CRESST Collaboration],
  Eur.\ Phys.\ J.\ C {\bf 76} (2016) no.1,  25
  [arXiv:1509.01515 [astro-ph.CO]].

\bibitem{edelweiss} S.~Scorza [EDELWEISS Collaboration],
  J.\ Phys.\ Conf.\ Ser.\  {\bf 718} (2016) no.4,  042053.

\bibitem{zeplin} V.~A.~Kudryavtsev [ZEPLIN-II Collaboration],
  J.\ Phys.\ Conf.\ Ser.\  {\bf 120} (2008) 042016.

\bibitem{ilidio1} I.~P.~Lopes, J.~Silk and S.~H.~Hansen,
  Mon.\ Not.\ Roy.\ Astron.\ Soc.\  {\bf 331} (2002) 361
  [astro-ph/0111530].
  
\bibitem{ilidio2} J.~Casanellas and I.~Lopes,
  Astrophys.\ J.\  {\bf 705} (2009) 135
  [arXiv:0909.1971 [astro-ph.CO]].
  
\bibitem{ilidio3} I.~Lopes, K.~Kadota and J.~Silk,
Astrophys.\ J.\ Lett.\  {\bf 780} (2014) L15
[arXiv:1310.0673 [astro-ph.SR]].

\bibitem{Vincent:2015gqa}
A.~C.~Vincent, A.~Serenelli and P.~Scott,
JCAP {\bf 1508} (2015) no.08,  040;
[arXiv:1504.04378 [hep-ph]].

\bibitem{Garani:2017jcj}
 R.~Garani and S.~Palomares-Ruiz,
JCAP {\bf 1705} (2017) no.05,  007;
[arXiv:1702.02768 [hep-ph]].

\bibitem{ilidio4} I.~Lopes, P.~Panci and J.~Silk,
Astrophys.\ J.\  {\bf 795} (2014) 162
[arXiv:1402.0682 [astro-ph.SR]].

\bibitem{ilidio5} I.~Lopes and J.~Silk,
Astrophys.\ J.\  {\bf 786} (2014) 25
[arXiv:1404.3909 [astro-ph.CO]].

\bibitem{ilidio6} J.~Lopes and I.~Lopes,
Astrophys.\ J.\  {\bf 827} (2016) no.2,  130
[arXiv:1607.08672 [astro-ph.CO]].

\bibitem{ilidio0} I.~Lopes,
Journal of Physics: Conference Series, Volume 665, Issue 1,
  article id. 012079 (2016)
[arXiv:1701.03926 [astro-ph.CO]].

\bibitem{kouvaris0} C.~Kouvaris and P.~Tinyakov,
  Phys.\ Rev.\ D {\bf 82} (2010) 063531
  [arXiv:1004.0586 [astro-ph.GA]].
  
\bibitem{kouvaris1} C.~Kouvaris and P.~Tinyakov,
  Phys.\ Rev.\ D {\bf 83} (2011) 083512
  [arXiv:1012.2039 [astro-ph.HE]].
  
\bibitem{kouvaris2} C.~Kouvaris and P.~Tinyakov,
Phys.\ Rev.\ Lett.\  {\bf 107} (2011) 091301
[arXiv:1104.0382 [astro-ph.CO]].

\bibitem{GPIL1} G.~Panotopoulos and I.~Lopes,
  Phys.\ Rev.\ D {\bf 96} (2017) no.2,  023016
  [arXiv:1707.06042 [hep-ph]].

\bibitem{GPIL2} G.~Panotopoulos and I.~Lopes,
  Phys.\ Rev.\ D {\bf 96} (2017) no.6,  063003
  [arXiv:1709.02272 [hep-ph]].
  
\bibitem{M4} L.~R.~Bedin et al., Astrophys. \ J. \ {\bf 697} (2009) 965.

\bibitem{NS1} C.~Kouvaris and P.~Tinyakov,
Phys. Rev. D \textbf{82} (2010), 063531
[arXiv:1004.0586 [astro-ph.GA]].

\bibitem{NS2} C.~Kouvaris,
Adv. High Energy Phys. \textbf{2013} (2013), 856196
[arXiv:1308.3222 [astro-ph.HE]].

\bibitem{NS3} G.~Panotopoulos and I.~Lopes,
Phys. Rev. D \textbf{96} (2017) no.2, 023016
[arXiv:1707.06042 [hep-ph]].

\bibitem{NS4} G.~Panotopoulos and I.~Lopes,
Phys. Rev. D \textbf{96} (2017) no.8, 083004
[arXiv:1709.06312 [hep-ph]].

\bibitem{NS5} A.~Das, T.~Malik and A.~C.~Nayak,
Phys. Rev. D \textbf{99} (2019) no.4, 043016
[arXiv:1807.10013 [hep-ph]].

\bibitem{NS6} D.~A.~Camargo, F.~S.~Queiroz and R.~Sturani,
JCAP \textbf{09} (2019), 051
[arXiv:1901.05474 [hep-ph]].

\bibitem{NS7} A.~Quddus, G.~Panotopoulos, B.~Kumar, S.~Ahmad and S.~Patra,
[arXiv:1902.00929 [nucl-th]].

\bibitem{NS8} H.~C.~Das, A.~Kumar, B.~Kumar, S.~Kumar Biswal, T.~Nakatsukasa, A.~Li and S.~Patra,
[arXiv:2002.00594 [nucl-th]].

\bibitem{guth} A.~H.~Guth,
  Phys.\ Rev.\ D {\bf 23} (1981) 347.
  
\bibitem{ref1} A.~H.~C{\'o}rsico, A.~D.~Romero, L.~G.~Althaus and J.~J.~Hermes, "The seismic properties of low-mass
He-core white dwarf stars," arXiv:1209.00613 [astro-ph.SR].

\bibitem{ref2} Y.~H.~Chen, "Asteroseismology of the DBV star CBS 114," arXiv:1604.5107 [astro-ph.SR].

\bibitem{ref3} G.~Fontain, P.~Brassard and P.~Bergeron, 2001, PASP, 113, 409.

\bibitem{ref4} A.~Bischoff-Kim and T.~S.~Metcalfe, 2011, MNRAS, 414, 404.

\bibitem{set} M.~McCullough and M.~Fairbairn,
  Phys.\ Rev.\ D {\bf 81} (2010) 083520
  [arXiv:1001.2737 [hep-ph]].
  
\bibitem{base2} M.~Cerme{\~n}o and M.~A.~P{\'e}rez-Garc{\'i}a,
  Phys.\ Rev.\ D {\bf 98} (2018) no.6,  063002
  [arXiv:1807.03318 [hep-ph]].
  
\bibitem{textbook} R.~Kippenhahn, A.~Weigert and A.~Weiss, 
\textit{Stellar structure and evolution}, Springer 2012.

\bibitem{chandra} S.~Chandrasekhar,
  Mon.\ Not.\ Roy.\ Astron.\ Soc.\  {\bf 95} (1935) 207.

\bibitem{model1} S.~L.~Shapiro and S.~A.~Teukolsky,
  \textit{Black holes, white dwarfs, and neutron stars: The physics of compact objects}, New York, USA: Wiley (1983) 645 p.
  
\bibitem{model2} D.~Koester and G.~Chanmugam 1990 
Rep. \ Prog. \ Phys. {\bf 53}, 837.

\bibitem{saltas} I.~D.~Saltas, I.~Sawicki and I.~Lopes,
  JCAP {\bf 1805} (2018) no.05,  028
  [arXiv:1803.00541 [astro-ph.CO]].
  
\bibitem{dearborn1} D.~Dearborn, G.~Raffelt, P.~Salati, J.~Silk and A.~Bouquet,
ApJ {\bf 354} (1990) 568.
  
\bibitem{dearborn2} D.~Dearborn, K.~Griest and G.~Raffelt, ApJ {\bf 368} (1991) 626.

\bibitem{kaplan} J.~Kaplan, F.~Martin de Volnay, C.~Tao and S.~Turck-Chieze,
ApJ {\bf 378} (1991) 315.
  
\bibitem{gould1} A.~Gould,
  Astrophys.\ J.\  {\bf 321} (1987) 560.
  
\bibitem{gould2} A.~Gould,
  Astrophys.\ J.\  {\bf 356} (1990) 302.
  
\bibitem{kouvaris3} C.~Kouvaris,
  Phys.\ Rev.\ D {\bf 92} (2015) no.7,  075001
  [arXiv:1506.04316 [hep-ph]].  
  
\bibitem{Griest:1986yu} K.~Griest and D.~Seckel,
  Nucl.\ Phys.\ B {\bf 283} (1987) 681
  Erratum: [Nucl.\ Phys.\ B {\bf 296} (1988) 1034].
  
\bibitem{ilidio7} I.~Lopes and J.~Silk,
  Astrophys.\ J.\  {\bf 757} (2012) 130.
  [arXiv: 1209.3631 [astro-ph.CO]].
  
\bibitem{Planck2015} P.~A.~R.~Ade {\it et al.} [Planck Collaboration],
Astron.\ Astrophys.\  {\bf 594} (2016) A13 [arXiv:1502.01589 [astro-ph.CO]].  
  
\bibitem{Planck2018} N.~Aghanim {\it et al.} [Planck Collaboration],
  arXiv:1807.06209 [astro-ph.CO].  
  
\bibitem{NFW} J.~F.~Navarro, C.~S.~Frenk and S.~D.~M.~White,
  Astrophys.\ J.\  {\bf 462} (1996) 563
  [astro-ph/9508025].
  
\bibitem{cirelli} G.~Bertone, M.~Cirelli, A.~Strumia and M.~Taoso,
  JCAP {\bf 0903} (2009) 009
  [arXiv:0811.3744 [astro-ph]].    
  
\bibitem{Catena:2009mf} R.~Catena and P.~Ullio,
  JCAP {\bf 1008} (2010) 004;
  [arXiv:0907.0018 [astro-ph.CO]].

\bibitem{salati} P.~Salati and J.~Silk, ApJ {\bf 338} (1989) 24.

\bibitem{Lopes:2014xaa} I.~Lopes and J.~Silk,
Astrophys.\ J.\  {\bf 786} (2014) 25;
[arXiv:1404.3909 [astro-ph.CO]].

\bibitem{singlet1} J.~McDonald,
  Phys.\ Rev.\ D {\bf 50} (1994) 3637
  [hep-ph/0702143 [HEP-PH]].
  
\bibitem{singlet2} S.~Andreas, T.~Hambye and M.~H.~G.~Tytgat,
  JCAP {\bf 0810} (2008) 034
  [arXiv:0808.0255 [hep-ph]].
  
\bibitem{singlet3} J.~M.~Cline, K.~Kainulainen, P.~Scott and C.~Weniger,
  Phys.\ Rev.\ D {\bf 88} (2013) 055025
   Erratum: [Phys.\ Rev.\ D {\bf 92} (2015) no.3,  039906]
  [arXiv:1306.4710 [hep-ph]].
  
\bibitem{singlet4} J.~A.~Casas, D.~G.~Cerdeño, J.~M.~Moreno and J.~Quilis,
  arXiv:1701.08134 [hep-ph].
  
\bibitem{inert1} L.~Lopez Honorez, E.~Nezri, J.~F.~Oliver and M.~H.~G.~Tytgat,
  JCAP {\bf 0702} (2007) 028
  [hep-ph/0612275].
  
\bibitem{inert2} L.~Lopez Honorez and C.~E.~Yaguna,
  JHEP {\bf 1009} (2010) 046
  [arXiv:1003.3125 [hep-ph]].
  
\bibitem{inert3} L.~Lopez Honorez and C.~E.~Yaguna,
  JCAP {\bf 1101} (2011) 002
  [arXiv:1011.1411 [hep-ph]].

\bibitem{Hportal} C.~Kouvaris, I.~M.~Shoemaker and K.~Tuominen,
  Phys.\ Rev.\ D {\bf 91} (2015) no.4,  043519
  [arXiv:1411.3730 [hep-ph]].
  
\bibitem{PDG} M.~Tanabashi {\it et al.} [Particle Data Group],
  Phys.\ Rev.\ D {\bf 98} (2018) no.3, 030001.  
  
\bibitem{age} B.~M.~S.~Hansen {\it et al.},
  Astrophys.\ J.\  {\bf 574} (2002) L155
  [astro-ph/0205087].
\end{thebibliography}
\end{document}